\RequirePackage[T1]{fontenc}
\documentclass[conference,a4paper]{IEEEtran}
\usepackage{amsmath,array,cite,graphicx}
\usepackage{tikz}
\usetikzlibrary{arrows.meta,positioning,shapes.geometric}
\providecommand{\FigureCoordinateScale}{1}
\newcommand{\FigureTextFont}{\fontsize{8}{9.6}\selectfont}
\newcommand{\FigurePanelFont}{\fontsize{8}{9.6}\selectfont}
\newsavebox{\FigureNaturalBox}
\newcommand{\FigureSizedInput}[2]{%
  \begingroup
  \def\FigureCoordinateScale{1}%
  \sbox{\FigureNaturalBox}{\input{#2}}%
  \pgfmathsetmacro{\FigureCoordinateScale}{(#1)/(\wd\FigureNaturalBox)}%
  \input{#2}%
  \endgroup
}

\graphicspath{{./}}
\begin{document}
\raggedbottom
\title{Directional and Quasi-Omnidirectional Indoor\\
Channel Measurements at Upper 6 GHz}
\author{\IEEEauthorblockN{Chengsong Fu\IEEEauthorrefmark{1}\IEEEauthorrefmark{2},
Hongbo Zhao\IEEEauthorrefmark{1}, Jianer Zhou\IEEEauthorrefmark{2},
Xiongbin Yu\IEEEauthorrefmark{2}\\
Weichao Li\IEEEauthorrefmark{2}, Chen Zhuang\IEEEauthorrefmark{1},
Meng Li\IEEEauthorrefmark{1}\IEEEauthorrefmark{2},
Wenquan Feng\IEEEauthorrefmark{1} and Xiaofeng Tao\IEEEauthorrefmark{3}}
\IEEEauthorblockA{\IEEEauthorrefmark{1}Beihang University, Beijing, China, bhzhb@buaa.edu.cn}
\IEEEauthorblockA{\IEEEauthorrefmark{2}Pengcheng Laboratory, Shenzhen, China, zhouje1005@gmail.com}
\IEEEauthorblockA{\IEEEauthorrefmark{3}Beijing University of Posts and Telecommunications, Beijing, China}
\IEEEauthorblockA{Corresponding authors: Hongbo Zhao, Jianer Zhou}}
\maketitle

\begin{abstract}
This paper presents wideband indoor channel measurements and
measurement-based statistical modeling of channel parameters in the upper
6 GHz (U6G) band using directional and quasi-omnidirectional antenna pairs.
Measurements span 6.425--7.125 GHz at 32 matched positions per configuration
in an open hall and an L-shaped corridor. System calibration and
frequency-dependent peak-gain normalization retain antenna radiation-pattern
weighting. Close-in and floating-intercept models are fitted to line-of-sight
(LOS) and non-LOS (NLOS) path loss, respectively. Shadow-fading residuals,
RMS delay spread (DS), and equivalent $K$ factors derived from power delay
profiles are characterized. LOS exponents range from 1.98 to 2.14. Despite
similar attenuation slopes, directional links exhibit smaller LOS residuals
and DS and higher equivalent $K$ factors. Hall DS medians are 1.78 and
36.52 ns for directional and quasi-omnidirectional links. Corridor NLOS
slopes are $-1.12$ and 2.54, respectively, while DS ordering varies by
position. These results characterize antenna-dependent parameters for U6G
indoor channels.
\end{abstract}

\begin{IEEEkeywords}
Upper 6 GHz, indoor propagation, antenna configuration, path loss, delay
spread, equivalent K factor.
\end{IEEEkeywords}

\section{Introduction}
The upper 6 GHz (U6G) band has attracted substantial interest as a mid-band
spectrum resource for next-generation mobile communications \cite{gsma6ghz}.
Directional and quasi-omnidirectional antennas represent directionally
concentrated and broad azimuthal coverage, respectively. Their radiation
patterns weight multipath contributions, affecting both attenuation and
delay-domain statistics observed in practical links. This weighting remains
after frequency-dependent boresight peak-gain normalization, motivating
wideband indoor measurements and joint statistical characterization of
U6G channel parameters.

Dabin \emph{et al.}\ \cite{dabin} compared antenna combinations in
2--6-GHz indoor LOS channels, while industrial LOS measurements examined
directional/omnidirectional reception at 4.145 GHz \cite{cebecioglu}.
Omnidirectional corridor measurements characterized path loss, DS, and
Ricean $K$ at 2.4, 5, and 6 GHz \cite{zhou}. Other studies characterized
LOS/NLOS corridors using 7--20-GHz narrowband measurements
\cite{moraitis} and extracted path loss, DS, and $K$ from 254 industrial
links using 10--12-GHz omnidirectional measurements, with comparison to
the 3GPP Indoor Factory model \cite{runtti}.

Angular scans at 6.75 and 16.95 GHz yielded directional statistics and
synthesized omnidirectional power delay profiles (PDPs) \cite{shakya}.
D-band LOS/NLOS office-corridor measurements showed increasing observable
path counts and DS with beamwidth \cite{kokkoniemi}. Geometry-related
studies characterized corridor LOS/NLOS path loss at 26, 32, and 39 GHz
\cite{pimienta} and validated an 8-GHz multi-ray loss model in an L-shaped
corridor \cite{yu}. Together, these findings motivate considering antenna
configuration and geometry jointly. Physical antenna-pair measurements
complement angular-scan synthesis by characterizing channels under a
specified pointing rule.

This work presents U6G measurements and statistical modeling of channel
parameters using physical directional and quasi-omnidirectional antenna
pairs at 6.425--7.125 GHz. The single-input single-output (SISO) responses
retain antenna-pattern weighting after system calibration and measured
peak-gain normalization. The contributions are
threefold. (i) Measurements cover 32 surveyed positions per configuration:
16 hall LOS, seven corridor LOS, and nine post-corner NLOS links, with
100 frequency-domain records per position under reproducible pointing
rules. (ii) Close-in (CI) and floating-intercept (FI) models characterize
LOS and NLOS path loss, respectively, alongside shadow-fading residuals
and descriptive distribution fits for DS and PDP-equivalent $K$.
(iii) Position-matched comparisons examine contrasting temporal dispersion
despite similar LOS slopes and a possible combined influence of geometry
and antenna-pattern weighting on the directional NLOS trend. Consecutive-day
measurements do not isolate antenna effects from temporal variation.

\section{Measurement and Parameter Extraction}
\subsection{Sounder, Antennas, and Environments}
The sounder combined an R\&S ZNA67 vector network analyzer (VNA), an
electro-optical transmitter, approximately 80 m of optical fiber, and a
remote RF front end to record complex $S_{21}(f)$ (Table~\ref{tab:setup}).
Figure~\ref{fig:system} shows the measurement photograph and system architecture.
At each position, $R=100$ complete frequency records were acquired at nominal
15-s intervals. The analyzed bandwidth $B=700$ MHz gives nominal delay
resolution $1/B=1.43$ ns; zero-padding only interpolates the delay grid.

\begin{table}[t]
\caption{Measurement and antenna configuration.}
\label{tab:setup}
\centering\footnotesize
\setlength{\tabcolsep}{3pt}\renewcommand{\arraystretch}{1.06}
\begin{tabular}{p{0.36\columnwidth}p{0.58\columnwidth}}
\hline
Parameter & Setting\\\hline
Sounder / source setting & R\&S ZNA67 / $-10$ dBm\\
Acquired frequencies & 6.0--7.5 GHz, 3001 points\\
Analyzed frequencies & 6.425--7.125 GHz, 1401 points\\
Bandwidth / spacing & 700 MHz / 500 kHz\\
Repeated acquisitions & $R=100$ per position, nominal 15-s interval\\
Tx / Rx height & 1.5 m / 1.5 m\\
Directional pair & TL-137SHAS20-T1, vertical pol.\\
Directional boresight peak gain & 20.10--20.71 dBi per antenna\\
Nominal H / V HPBW & $29$--$35^{\circ}$ / $23$--$29^{\circ}$\\
Quasi-omnidirectional pair & UWB2, linear polarization\\
Quasi-omni boresight peak gain & 4.33--5.87 dBi per antenna\\
Hall positions & 16 LOS, 7.7--32.0 m\\
Corridor positions & 7 LOS at 6--18 m, 9 NLOS at 20--36 m\\\hline
\end{tabular}
\par\smallskip\raggedright\scriptsize
\end{table}

\begin{figure*}[t]
\centering
\input{measurement_setup.tex}
\caption{Measurement system: (a) photograph of the VNA-based sounder and
tripod-mounted directional antenna pair, and (b) system architecture
with an optical fiber link and remote RF front end.}
\label{fig:system}
\end{figure*}

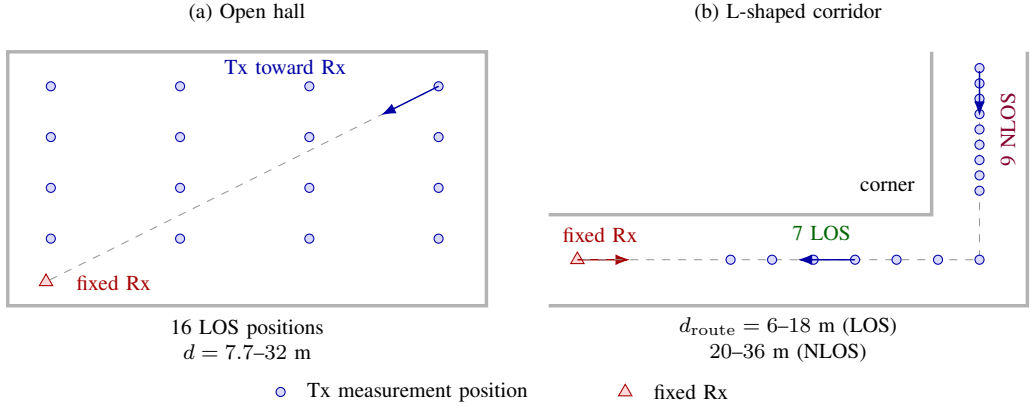
\begin{figure*}[t]
\centering
  \begingroup
  \def\FigureCoordinateScale{1}%
  \sbox{\FigureNaturalBox}{% Two scenario panels with the original positions and pointing rules.
\begin{tikzpicture}[
scale=\FigureCoordinateScale,
font=\FigureTextFont,
title/.style={font=\FigurePanelFont},
tx/.style={circle,draw=blue!65!black,fill=blue!15,minimum size=3.4pt,inner sep=0pt},
rx/.style={regular polygon,regular polygon sides=3,draw=red!65!black,
fill=red!15,minimum size=5.8pt,inner sep=0pt},
wall/.style={line width=1.3pt,gray!60},
beam/.style={-Latex,line width=.55pt,blue!65!black}]
\path[use as bounding box] (-.15,-1.05) rectangle (10.90,3.30);

% (a) Open hall: all 16 measured LOS positions are retained.
\begin{scope}
\node[title] at (2.5,3.05) {(a) Open hall};
\draw[wall] (0,0) rectangle (5,2.65);
\node[rx] (hallrx) at (.4,.25) {};
\node[anchor=west,text=red!65!black] at (.62,.24) {fixed Rx};
\foreach \x in {.45,1.8,3.15,4.5}
  \foreach \y in {.70,1.23,1.76,2.29}
    \node[tx] at (\x,\y) {};
\draw[dashed,black!40,line width=.4pt] (hallrx)--(4.5,2.29);
\draw[beam] (4.5,2.29)--(3.9,1.99);
\node[inner sep=0pt,text=blue!65!black] at (2.9,2.49) {Tx toward Rx};
\node[align=center] at (2.5,-.35) {16 LOS positions\\$d=7.7$--$32$ m};
\end{scope}

% (b) Corridor: seven LOS and nine NLOS positions and the original pointing rule.
\begin{scope}[xshift=5.65cm]
\node[title] at (2.5,3.05) {(b) L-shaped corridor};
\draw[wall] (0,0)--(5,0)--(5,2.65);
\draw[wall] (0,.95)--(4,.95)--(4,2.65);
\node[rx] (corrrx) at (.3,.48) {};
\node[anchor=south west,text=red!65!black] at (.05,.56) {fixed Rx};
\draw[-Latex,red!65!black,line width=.55pt] (.3,.48)--(.85,.48);
\draw[dashed,black!40,line width=.4pt] (corrrx)--(4.5,.48)--(4.5,2.48);
\foreach \x in {1.90,2.3333,2.7667,3.20,3.6333,4.0667,4.50}
  \node[tx] at (\x,.48) {};
\foreach \y in {1.20,1.36,1.52,1.68,1.84,2.00,2.16,2.32,2.48}
  \node[tx] at (4.5,\y) {};
\draw[beam] (3.20,.48)--(2.60,.48);
\draw[beam] (4.5,2.45)--(4.5,1.98);
\node[text=green!40!black] at (2.85,.76) {7 LOS};
\node[rotate=90,text=purple!70!black] at (4.80,1.84) {9 NLOS};
\node[anchor=south east] at (3.92,1.10) {corner};
\node[align=center] at (2.5,-.35)
  {$d_{\rm route}=6$--$18$ m (LOS)\\$20$--$36$ m (NLOS)};
\end{scope}

% One compact shared legend centered below the two panels.
\node[tx] at (2.85,-.9) {};
\node[anchor=west] at (3.03,-.9) {Tx measurement position};
\node[rx] at (6.45,-.9) {};
\node[anchor=west] at (6.65,-.9) {fixed Rx};
\end{tikzpicture}}%
  \pgfmathsetmacro{\FigureCoordinateScale}{(0.78\textwidth)/(\wd\FigureNaturalBox)}%
  % Two scenario panels with the original positions and pointing rules.
\begin{tikzpicture}[
scale=\FigureCoordinateScale,
font=\FigureTextFont,
title/.style={font=\FigurePanelFont},
tx/.style={circle,draw=blue!65!black,fill=blue!15,minimum size=3.4pt,inner sep=0pt},
rx/.style={regular polygon,regular polygon sides=3,draw=red!65!black,
fill=red!15,minimum size=5.8pt,inner sep=0pt},
wall/.style={line width=1.3pt,gray!60},
beam/.style={-Latex,line width=.55pt,blue!65!black}]
\path[use as bounding box] (-.15,-1.05) rectangle (10.90,3.30);

% (a) Open hall: all 16 measured LOS positions are retained.
\begin{scope}
\node[title] at (2.5,3.05) {(a) Open hall};
\draw[wall] (0,0) rectangle (5,2.65);
\node[rx] (hallrx) at (.4,.25) {};
\node[anchor=west,text=red!65!black] at (.62,.24) {fixed Rx};
\foreach \x in {.45,1.8,3.15,4.5}
  \foreach \y in {.70,1.23,1.76,2.29}
    \node[tx] at (\x,\y) {};
\draw[dashed,black!40,line width=.4pt] (hallrx)--(4.5,2.29);
\draw[beam] (4.5,2.29)--(3.9,1.99);
\node[inner sep=0pt,text=blue!65!black] at (2.9,2.49) {Tx toward Rx};
\node[align=center] at (2.5,-.35) {16 LOS positions\\$d=7.7$--$32$ m};
\end{scope}

% (b) Corridor: seven LOS and nine NLOS positions and the original pointing rule.
\begin{scope}[xshift=5.65cm]
\node[title] at (2.5,3.05) {(b) L-shaped corridor};
\draw[wall] (0,0)--(5,0)--(5,2.65);
\draw[wall] (0,.95)--(4,.95)--(4,2.65);
\node[rx] (corrrx) at (.3,.48) {};
\node[anchor=south west,text=red!65!black] at (.05,.56) {fixed Rx};
\draw[-Latex,red!65!black,line width=.55pt] (.3,.48)--(.85,.48);
\draw[dashed,black!40,line width=.4pt] (corrrx)--(4.5,.48)--(4.5,2.48);
\foreach \x in {1.90,2.3333,2.7667,3.20,3.6333,4.0667,4.50}
  \node[tx] at (\x,.48) {};
\foreach \y in {1.20,1.36,1.52,1.68,1.84,2.00,2.16,2.32,2.48}
  \node[tx] at (4.5,\y) {};
\draw[beam] (3.20,.48)--(2.60,.48);
\draw[beam] (4.5,2.45)--(4.5,1.98);
\node[text=green!40!black] at (2.85,.76) {7 LOS};
\node[rotate=90,text=purple!70!black] at (4.80,1.84) {9 NLOS};
\node[anchor=south east] at (3.92,1.10) {corner};
\node[align=center] at (2.5,-.35)
  {$d_{\rm route}=6$--$18$ m (LOS)\\$20$--$36$ m (NLOS)};
\end{scope}

% One compact shared legend centered below the two panels.
\node[tx] at (2.85,-.9) {};
\node[anchor=west] at (3.03,-.9) {Tx measurement position};
\node[rx] at (6.45,-.9) {};
\node[anchor=west] at (6.65,-.9) {fixed Rx};
\end{tikzpicture}%
  \endgroup

\caption{Measurement layouts: (a) open hall with 16 LOS positions and
(b) L-shaped
corridor with seven LOS and nine NLOS positions. Schematics are not to scale.
LOS boresights are mutually aligned. After the corner, Tx points along the
NLOS trajectory toward the corner and the Rx orientation remains fixed.}
\label{fig:sites}
\end{figure*}

The hall has 16 LOS positions; the L-shaped corridor has seven LOS and
nine NLOS positions (Fig.~\ref{fig:sites}). A 0.01-m-accuracy laser
rangefinder established coordinates for the hall Euclidean separation $d$
and LOS boresight azimuths. Corridor distance $d_{\rm route}$ follows the
surveyed trajectory. Both configurations reused the survey marks on
consecutive days with unchanged geometry and static layout, although
day-to-day variation remains possible.

\begin{figure}[t]
\centering
\includegraphics[width=0.97\columnwidth]{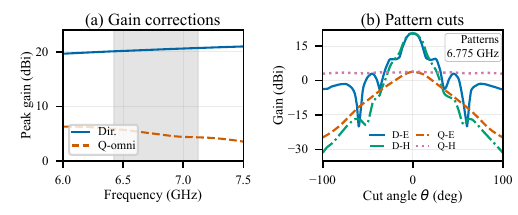}
\caption{Antenna characterization: (a) measured boresight peak gains versus
frequency used for peak-gain normalization (shading: analyzed band);
(b) manufacturer and anechoic-chamber E/H-plane cuts at 6.775 GHz,
distinguished by color and line style.}
\label{fig:antennas}
\end{figure}

LOS boresights were aligned using coordinate-derived azimuths and tripod
angular scales. In corridor NLOS, Rx remained directed along the initial
segment while Tx pointed along the trajectory toward the corner, probing
reflected and diffracted paths. ``Directional'' denotes this reproducible
pointing rule, without angular sweeps or turntable-based synthesis. The
physical UWB2 pair provides quasi-omnidirectional azimuthal coverage.
Figure~\ref{fig:antennas} shows frequency-dependent boresight peak gains and
E/H-plane antenna-pattern cuts from manufacturer characterization and our
anechoic-chamber measurements.

\subsection{Calibration and Parameter Definition}
A campaign-specific direct-connection reference provides $H_{\rm sys}(f)$.
Neglecting antenna group-delay variation, calibration removes the system
response and frequency-dependent boresight peak gains while retaining the
normalized field-amplitude patterns:
\begin{equation}
 H_{{\rm eff},r}(f)=\frac{H_{{\rm meas},r}(f)}{H_{\rm sys}(f)}
 10^{-[G_{\rm t}^{\rm pk}(f)+G_{\rm r}^{\rm pk}(f)]/20},
\label{eq:cal}
\end{equation}
where $H_{{\rm meas},r}(f)$ is complex $S_{21}$ for record $r=1,\ldots,R$
at frequency $f$, and $G_{\rm t}^{\rm pk}(f)$ and $G_{\rm r}^{\rm pk}(f)$
are measured Tx/Rx boresight peak gains in dBi. Antenna-pattern weighting
remains:
\begin{equation}
\begin{aligned}
 H_{{\rm eff},r}(f)={}&\sum_{\ell=1}^{L}\alpha_{\ell,r}(f)
 A_{\rm t}(f,\boldsymbol{\Omega}_{{\rm t},\ell})\\
 &\quad\times A_{\rm r}(f,\boldsymbol{\Omega}_{{\rm r},\ell})
 e^{-j2\pi f\tau_{\ell,r}},
\end{aligned}
\label{eq:pattern_weighting}
\end{equation}
Here, $A_{\rm t}$ and $A_{\rm r}$ are normalized Tx/Rx field-amplitude
patterns dependent on frequency and direction. For multipath index
$\ell=1,\ldots,L$ and record $r$, $\alpha_{\ell,r}(f)$ and $\tau_{\ell,r}$
are the complex propagation coefficient and delay;
$\boldsymbol{\Omega}_{{\rm t},\ell}$ and $\boldsymbol{\Omega}_{{\rm r},\ell}$
are departure and arrival directions in antenna coordinates, and $j$ is the
imaginary unit.

\begin{table*}[t]
\caption{Measured path-loss and delay-domain descriptors with 3GPP Indoor-Office reference values at $f_{\rm c}=6.775$ GHz.}
\label{tab:parameters}
\centering\footnotesize
\setlength{\tabcolsep}{4.0pt}\renewcommand{\arraystretch}{1.08}
\begin{tabular}{lllclrrrrr}
\hline
Scenario & Condition & Configuration & $N_{\rm pos}$ & Model & $PL_0$ or $\alpha$ & $n$ or $\beta$ & $\sigma_{\rm SF}$ & Median $\tau_{\rm RMS}$ & $K$ value$^{c}$\\
 & & & & & (dB) & & (dB) & (ns) & (dB)\\\hline
Open hall & LOS & Directional & 16 & CI & 49.06$^{a}$ & 2.10 & 1.44 & 1.78 & 9.34\\
Open hall & LOS & Quasi-omni & 16 & CI & 49.06$^{a}$ & 2.11 & 2.60 & 36.52 & 2.30\\
L-corridor & LOS & Directional & 7 & CI & 49.06$^{a}$ & 1.98 & 0.23 & 3.91 & 9.38\\
L-corridor & LOS & Quasi-omni & 7 & CI & 49.06$^{a}$ & 2.14 & 1.33 & 15.16 & $-0.28$\\
L-corridor & NLOS & Directional & 9 & FI & 118.28 & $-1.12$ & 1.83 & 24.34 & 1.88\\
L-corridor & NLOS & Quasi-omni & 9 & FI & 53.91 & 2.54 & 3.01 & 30.08 & $-3.56$\\\hline
3GPP reference & LOS & Indoor-Office & -- & 3GPP & -- & 1.73$^{b}$ & 3.00 & 19.91 & 7.00\\
3GPP reference & NLOS & Indoor-Office & -- & 3GPP & -- & 3.83$^{b}$ & 8.03 & 37.81 & --\\\hline
\end{tabular}
\par\smallskip\begin{minipage}{0.98\textwidth}\scriptsize
$^{a}$The CI model fixes $PL_0$ to the 1-m FSPL at $f_{\rm c}$. The FI intercept $\alpha$ is a fitted extrapolation. Corridor fits use $d_{\rm route}$.
$^{b}$The 3GPP coefficients multiply $10\log_{10}d_{\rm 3D}$. The NLOS model takes the maximum of its LOS and NLOS branches.
$^{c}$Measured entries are medians of the PDP proxy $K_{\rm eq}$. The 3GPP entry is the mean Ricean $K$ and uses a different estimator.
Measured $\sigma_{\rm SF}$ is the RMS of the fitted dB residuals. 3GPP gives ensemble shadow fading. $\tau_{\rm RMS}$ uses the stated 200-ns window.
\end{minipage}
\end{table*}

\begin{figure*}[t]
\centering\includegraphics[width=\textwidth]{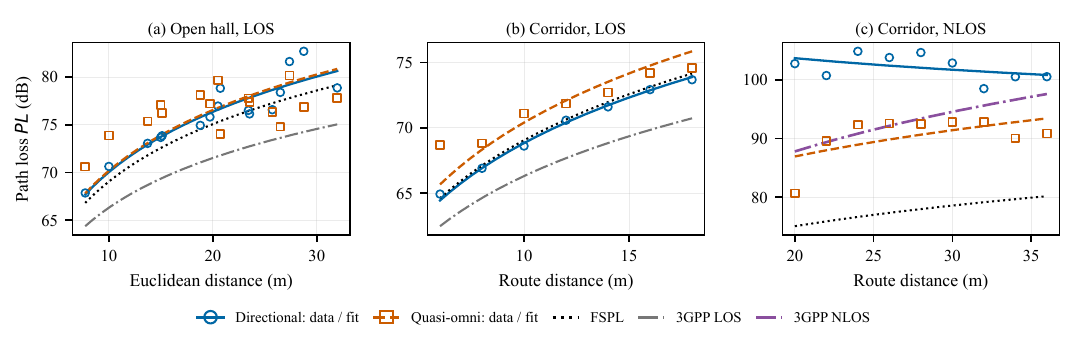}
\caption{Peak-gain-normalized PL, LOS CI and NLOS FI fits, FSPL, and 3GPP
Indoor-Office LOS and NLOS references at 6.775 GHz. Corridor fits use route
distance $d_{\rm route}$, which also serves as a proxy for $d_{\rm 3D}$ in
the illustrative 3GPP reference curves.}
\label{fig:loss}
\end{figure*}

The antenna-filtered path loss (PL) is
\begin{equation}
 PL=-10\log_{10}\left\langle|H_{{\rm eff},r}(f)|^2\right\rangle_{r,f}.
\label{eq:pathloss}
\end{equation}
The brackets average over records and analyzed frequencies. For
$N_{\rm pos}$ fitted positions, close-in (CI) and floating-intercept (FI)
models share
\begin{equation}
 \widehat{PL}(d)=A+10b\log_{10}(d/d_0),\qquad d_0=1~\mathrm m .
\label{eq:pl}
\end{equation}
Here, $\widehat{PL}(d)$ is modeled PL at distance $d$, $d_0$ is the reference
distance, and $A$ and $b$ are the intercept in dB and dimensionless slope.
LOS uses CI with
$(A,b)=(PL_0,n)$, fitting only $n$ and fixing
$PL_0=\mathrm{FSPL}(f_{\rm c},d_0)=49.06$ dB at $f_{\rm c}=6.775$ GHz.
Post-corner NLOS uses FI with both $(A,b)=(\alpha,\beta)$ fitted;
$\alpha$ is an extrapolated 1-m intercept. FSPL is
$PL_{\rm FS}(d)=PL_0+20\log_{10}(d/d_0)$. For measured $PL_i$ at distance
$d_i$, residual $X_i=PL_i-\widehat{PL}(d_i)$ gives
$\sigma_{\rm SF}=\sqrt{N_{\rm pos}^{-1}\sum_{i=1}^{N_{\rm pos}}X_i^2}$
for either model.

An RMS-normalized Hann window and 8192-point IFFT yield each PDP; the
position PDP is the binwise median over $R$ records. Noise is estimated
from 1.2--1.8 $\mu$s after upper-tail outlier rejection. Within the 200-ns
excess-delay window, bins exceeding the cleaned noise distribution's 99.9th
percentile plus 3 dB are retained. Their index set is $\mathcal S$, with
linear power $P_k$ and delay $\tau_k$ at bin $k$. Unlike multipath index
$\ell$, $k$ indexes discrete delay bins. The RMS delay spread (DS) is
\begin{equation}
 \tau_{\rm RMS}=\sqrt{\frac{\sum_{k\in\mathcal S}P_k(\tau_k-\bar\tau)^2}
 {\sum_{k\in\mathcal S}P_k}},\qquad
 \bar\tau=\frac{\sum_{k\in\mathcal S}P_k\tau_k}
 {\sum_{k\in\mathcal S}P_k}.
\label{eq:ds}
\end{equation}
Here, $\bar\tau$ is the retained bins' power-weighted mean delay. Total
power is $P_{\rm tot}=\sum_{k\in\mathcal S}P_k$; main-path power
$P_{\rm main}$ is integrated within $1/B$ on either side of the strongest
component. The remaining power $P_{\rm res}=P_{\rm tot}-P_{\rm main}$ gives
the PDP-equivalent $K$ factor in dB:
\begin{equation}
 K_{\rm eq}=10\log_{10}\frac{P_{\rm main}}{P_{\rm res}}
 =10\log_{10}\frac{P_{\rm main}}{P_{\rm tot}-P_{\rm main}}.
\label{eq:k}
\end{equation}
This PDP power-ratio proxy is not a fitted Ricean-amplitude parameter.

\begin{figure*}[t]
\centering\includegraphics[width=0.9\textwidth]{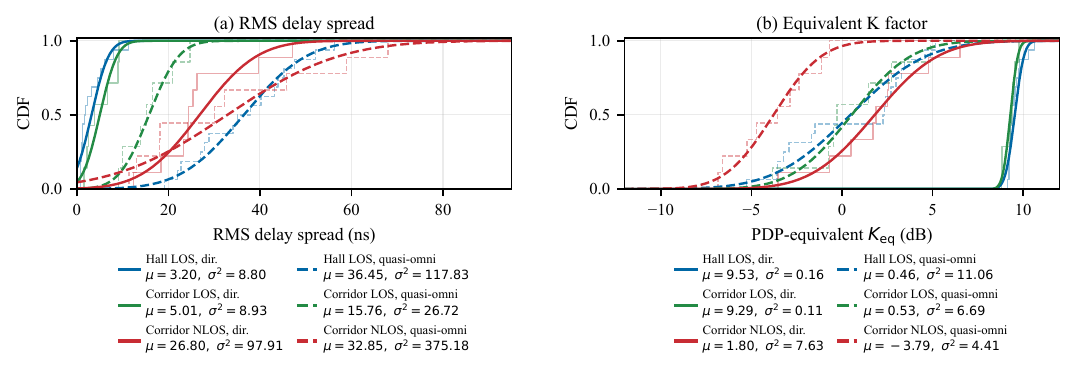}
\caption{Empirical position-level CDFs and descriptive Gaussian CDF fits on
linear axes. Solid and dashed traces denote directional and physical
quasi-omnidirectional configurations. Legends give the fitted mean $\mu$
and variance $\sigma^2$ for all six groups (ns and ns$^2$ for DS, dB and
dB$^2$ for $K_{\rm eq}$).}
\label{fig:cdf}
\end{figure*}

\section{Results and Discussion}
\subsection{Path Loss and Shadow-Fading Residuals}
Table~\ref{tab:parameters} and Fig.~\ref{fig:loss} show similar LOS slopes
but opposite corridor NLOS trends. FSPL and 3GPP Indoor-Office LOS and NLOS
provide references at 6.775 GHz \cite{3gpp}. The corridor reference curves
use $d_{\rm route}$ as a proxy for the model's direct three-dimensional
separation $d_{\rm 3D}$ and are therefore illustrative comparisons.

The CI exponents for directional and quasi-omnidirectional pairs are
2.10/2.11 in hall LOS and 1.98/2.14 in corridor LOS. Over the measured ranges
and configurations, these fitted exponents are close to the free-space
value of 2 and exceed the 3GPP LOS coefficient of 1.73.
Median excess losses $PL_i-PL_{\rm FS}(d_i)$ are
1.08/2.31 dB and $-0.21$/1.23 dB, respectively.

In corridor NLOS, the directional FI slope is $\beta=-1.12$. With Tx
pointing along the trajectory toward the corner and Rx fixed along the
initial segment (Fig.~\ref{fig:sites}), the horizontal Tx-to-Rx bearing is
more oblique to the Tx boresight at shorter post-corner distances. The
resulting Tx/Rx antenna-pattern weighting of reflected and diffracted paths
can impose angular attenuation whose reduction along the route outweighs
additional propagation loss. Position-dependent corner propagation
\cite{yu} and beam-pointing sensitivity in corridors \cite{diba} provide
related context for this negative local slope.

The quasi-omnidirectional pair, less sensitive to azimuthal misalignment,
yields $\beta=2.54$, qualitatively consistent with the increasing 3GPP NLOS
trend \cite{3gpp}. The 3GPP coefficient 3.83 is defined for
$10\log_{10}(d_{\rm 3D})$, whereas the measured FI slope is fitted against
$d_{\rm route}$. The plotted reference uses the route-distance proxy
specified above.

Directional\slash quasi-omnidirectional $\sigma_{\rm SF}$ values are 1.44/2.60 dB
in hall LOS, 0.23/1.33 dB in corridor LOS, and 1.83/3.01 dB in corridor NLOS.
The smaller directional residuals are consistent with off-axis suppression
but remain site-specific fits rather than 3GPP ensemble shadow fading.
Similar LOS PL slopes nevertheless coexist with different delay dispersion.

\subsection{RMS Delay Spread}
Position-level CDFs appear in Fig.~\ref{fig:cdf}.
Directional\slash quasi-omnidirectional DS medians are
1.78/36.52 ns in hall LOS and 3.91/15.16 ns in corridor LOS. Directional DS
is lower at all 16 hall and seven corridor LOS positions. The hall paired
difference (quasi-omnidirectional minus directional) has median 32.22 ns and
interquartile range (IQR) 24.81--41.56 ns. Directional antenna patterns
suppress off-axis delayed energy, whereas the quasi-omnidirectional pair
retains stronger late components.

Corridor NLOS medians are 24.34/30.08 ns, with directional DS lower at five
of nine positions and higher at four. The paired median is 5.74 ns with
IQR $-14.92$--33.42 ns (Table~\ref{tab:paired}). This mixed ordering reflects
position-dependent reflected and diffracted contributions under the fixed
pointing rule.

\subsection{PDP-Equivalent K Factor}
The $K_{\rm eq}$ medians for directional and quasi-omnidirectional pairs are
9.34/2.30 dB in hall LOS, 9.38/$-0.28$ dB in corridor LOS, and
1.88/$-3.56$ dB in corridor NLOS. Directional $K_{\rm eq}$ is higher at every
matched position. In LOS, its aligned main-path region holds a larger power
fraction, whereas the strong delayed components retained by the
quasi-omnidirectional pair increase $P_{\rm res}$
relative to $P_{\rm main}$, allowing negative $K_{\rm eq}$.

A value of $K_{\rm eq}$ close to 0 dB indicates that the power integrated
within the main-path region is comparable to the total retained power
outside that region. A negative value indicates that the latter exceeds
the former. Neither case implies the absence of a geometrical LOS path,
since this metric quantifies power concentration rather than LOS visibility.

DS and $K_{\rm eq}$ are complementary: delayed energy generally increases
the second delay moment and residual power, raising DS and lowering
$K_{\rm eq}$. Their group medians need not share an
ordering because DS also depends on path delays.

\subsection{Paired Comparison and Standard-Model Interpretation}
\begin{table}[t]
\caption{Position-matched differences: quasi-omni minus directional.}
\label{tab:paired}
\centering\footnotesize
\setlength{\tabcolsep}{3pt}\renewcommand{\arraystretch}{1.10}
\begin{tabular}{lrrr}
\hline
Statistic & Hall LOS & Corr. LOS & Corr. NLOS\\\hline
$N_{\rm pos}$ & 16 & 7 & 9\\
Median $\Delta PL$ (dB) & 1.32 & 1.28 & $-11.12$\\
Median $\Delta$DS (ns) & 32.22 & 9.89 & 5.74\\
$\Delta$DS IQR (ns) & [24.8,41.6] & [6.9,11.6] & [$-14.9$,33.4]\\
Positions with $\Delta$DS$>0$ & 16/16 & 7/7 & 5/9\\
Median $\Delta K_{\rm eq}$ (dB) & $-7.00$ & $-9.99$ & $-7.23$\\\hline
\end{tabular}
\par\smallskip\raggedright\scriptsize
Differences are formed before taking the median. They are not differences
of group medians. IQR denotes the 25th--75th percentiles. Separate-day
campaigns and spatial dependence preclude an independent causal interpretation.
\end{table}

The median paired difference $\Delta PL$ (quasi-omnidirectional minus
directional) is 1.32 dB in hall LOS, 1.28 dB in corridor LOS, and
$-11.12$ dB in corridor NLOS. After blockage, broader angular coverage
captures more reflected and diffracted power, lowering peak-gain-normalized
PL. This difference retains antenna-pattern weighting rather than
representing antenna-independent attenuation.

The 3GPP LOS and NLOS DS medians at 6.775 GHz are 19.91 and 37.81 ns,
respectively \cite{3gpp}, broadly comparable with the quasi-omnidirectional
results. These standardized values provide references rather than
measurement labels. The 3GPP LOS Ricean-$K$ mean and standard deviation are
7 and 4 dB, whereas Table~\ref{tab:parameters} reports median $K_{\rm eq}$
using the definition in (\ref{eq:k}).

\subsection{Sensitivity and Scope of the Evidence}
For 100--300-ns metric windows, quasi-omnidirectional median DS spans
11.81--64.09 ns in the hall and 12.05--69.30 ns in corridor NLOS; 12 of 16
hall PDPs reach the selected 200-ns boundary. All reported configurations
therefore use the same stated window and detector. Expanding the main-path
half-width from one to two resolution cells changes hall directional
$K_{\rm eq}$ from 9.34 to 26.49 dB. Median-PDP metrics also differ from
per-record aggregation. Lower quasi-omnidirectional boresight peak gain
raises the peak-gain-normalized noise floor; despite separately estimated
noise thresholds, unequal weak-path sensitivity contributes to PDP, DS,
and $K_{\rm eq}$ differences.

\section{Conclusion}
This work combines U6G wideband indoor measurements using directional and
quasi-omnidirectional antenna pairs with measurement-based path loss
modeling and statistical characterization of DS and PDP-equivalent $K$
factors. LOS CI exponents are near two, but DS differs markedly.
Directional links exhibit smaller LOS residuals and DS and higher
$K_{\rm eq}$ at every matched position. In corridor NLOS, fixed pointing
geometry offers an explanation for the negative local directional FI slope.
The quasi-omnidirectional slope is positive, while DS ordering varies by
position. Peak-gain normalization retains angular filtering and unequal
weak-path sensitivity. The results characterize physical antenna-pair
channels without angular-scan synthesis.

\section*{Acknowledgment}
This work was supported in part by Chinese National Natural Science Foundation of
China under Grant No. 62571023, and in part by the Major Key Project of PCL (PCL2025A02).

\bibliographystyle{IEEEtran}
\bibliography{references}

% Generated by IEEEtran.bst, version: 1.14 (2015/08/26)
\begin{thebibliography}{10}
\providecommand{\url}[1]{#1}
\csname url@samestyle\endcsname
\providecommand{\newblock}{\relax}
\providecommand{\bibinfo}[2]{#2}
\providecommand{\BIBentrySTDinterwordspacing}{\spaceskip=0pt\relax}
\providecommand{\BIBentryALTinterwordstretchfactor}{4}
\providecommand{\BIBentryALTinterwordspacing}{\spaceskip=\fontdimen2\font plus
\BIBentryALTinterwordstretchfactor\fontdimen3\font minus
  \fontdimen4\font\relax}
\providecommand{\BIBforeignlanguage}[2]{{%
\expandafter\ifx\csname l@#1\endcsname\relax
\typeout{** WARNING: IEEEtran.bst: No hyphenation pattern has been}%
\typeout{** loaded for the language `#1'. Using the pattern for}%
\typeout{** the default language instead.}%
\else
\language=\csname l@#1\endcsname
\fi
#2}}
\providecommand{\BIBdecl}{\relax}
\BIBdecl

\bibitem{gsma6ghz}
\BIBentryALTinterwordspacing
{GSMA}, ``6 {GHz} in europe: Mobile evolution in 6.425--7.125 {GHz},'' GSMA,
  Tech. Rep., Jun. 2025. [Online]. Available:
  \url{https://www.gsma.com/connectivity-for-good/spectrum/wp-content/uploads/2025/07/6-GHz-in-Europe.pdf}
\BIBentrySTDinterwordspacing

\bibitem{dabin}
J.~A. Dabin, A.~M. Haimovich, and H.~Grebel, ``A statistical ultra-wideband
  indoor channel model and the effects of antenna directivity on path loss and
  multipath propagation,'' \emph{IEEE J. Sel. Areas Commun.}, vol.~24, no.~4,
  pp. 752--758, Apr. 2006.

\bibitem{cebecioglu}
B.~B. Cebecioglu \emph{et~al.}, ``Sub-6 {GHz} channel modeling and evaluation
  in indoor industrial environments,'' \emph{IEEE Access}, vol.~10, pp.
  127\,742--127\,753, 2022.

\bibitem{zhou}
Z.~Zhou, L.~Zhang, X.~Chen, C.-X. Wang, and J.~Huang, ``Multi-frequency
  wireless channel measurements and characteristics analysis in indoor corridor
  scenarios,'' in \emph{Proc. IEEE 94th Veh. Technol. Conf. (VTC2021-Fall)},
  2021, pp. 1--5.

\bibitem{moraitis}
N.~Moraitis, A.~Rogaris, I.~Popescu, and K.~S. Nikita, ``Indoor corridor
  channel measurements and characterization in the upper mid-band,'' in
  \emph{Proc. 19th Eur. Conf. Antennas Propag. (EuCAP)}, 2025, pp. 1--5.

\bibitem{runtti}
J.-M. Runtti \emph{et~al.}, ``Mid-band propagation measurements in industrial
  environments,'' in \emph{Proc. 19th Eur. Conf. Antennas Propag. (EuCAP)},
  2025, pp. 1--5.

\bibitem{shakya}
D.~Shakya \emph{et~al.}, ``Propagation measurements and channel models in
  indoor environment at 6.75 {GHz} {FR1(C)} and 16.95 {GHz} {FR3} upper-mid
  band spectrum for {5G} and {6G},'' in \emph{Proc. IEEE Global Commun. Conf.
  (GLOBECOM)}, 2024, pp. 998--1003.

\bibitem{kokkoniemi}
J.~Kokkoniemi, V.~Hovinen, K.~Nevala, and M.~Juntti, ``Initial results on
  {D}-band channel measurements in {LOS} and {NLOS} office corridor
  environment,'' in \emph{Proc. 16th Eur. Conf. Antennas Propag. (EuCAP)},
  2022, pp. 1--5.

\bibitem{pimienta}
D.~P. del Valle, L.~Mendo, J.~M. Riera, and P.~G. del Pino, ``Path loss results
  in an indoor corridor scenario at the 26, 32 and 39 {GHz} millimeter-wave
  bands,'' in \emph{Proc. 15th Eur. Conf. Antennas Propag. (EuCAP)}, 2021, pp.
  1--5.

\bibitem{yu}
Y.~Yu, T.-T. Liu, Y.-Q. Bao, H.-F. Xu, S.~Jiang, and W.-H. Zeng, ``A {3D}
  multi-rays path loss model above 6 {GHz} under indoor environments with
  regular structures,'' \emph{Prog. Electromagn. Res. C}, vol. 126, pp.
  157--172, 2022.

\bibitem{3gpp}
{3GPP}, ``Study on channel model for frequencies from 0.5 to 100 {GHz},'' 3GPP,
  Tech. Rep. TR 38.901, Jul. 2026, version 19.4.0, Release 19.

\bibitem{diba}
F.~D. Diba, M.~A. Samad, and D.-Y. Choi, ``Centimeter and millimeter-wave
  propagation characteristics for indoor corridors: Results from measurements
  and models,'' \emph{IEEE Access}, vol.~9, pp. 158\,726--158\,737, 2021.

\end{thebibliography}
\end{document}